%% file: main.tex
\documentclass[%
 aip,
 jmp,%
 amsmath,amssymb,
 reprint,%
]{revtex4-1}

\usepackage[english]{babel}

\usepackage{braket}

\usepackage{amsmath, amssymb}
\usepackage{color}
\usepackage{graphicx}
\usepackage{graphics}

\usepackage{xcolor}

\usepackage{mathtools}
\usepackage{mathrsfs,amsmath}

\usepackage{pdfpages}

\newcommand{\vect}[1]{\boldsymbol{#1}}

\newcommand{\kS}{\kappa_\mathrm{S}}
\newcommand{\kA}{\kappa_\mathrm{A}}
\newcommand{\kB}{\kappa_\mathrm{B}}

\newcommand{\kBolt}{k_\mathrm{B}}

\newcommand{\alphaB}{\alpha_\mathrm{B}}

\newcommand{\betaB}{\beta_\mathrm{B}}

\newcommand{\TS}{T_\mathrm{S}}
\newcommand{\TB}{T_\mathrm{B}}

\newcommand{\zm}{z_\mathrm{m}}

\newcommand{\G}{\mathcal{G}}
\newcommand{\R}{\vect{r}}
\newcommand{\Intd}{\mathrm{d}}

\newcommand{\DeltaPara}{\Delta_{\parallel}}
\newcommand{\vt}{\tilde{v}}
\newcommand{\ut}{\tilde{u}}
\newcommand{\psit}{\tilde{\psi}}
\newcommand{\Gt}{\tilde{\G}}

\newcommand{\Qt}{\tilde{Q}}

\begin{document}

\preprint{AIP/123-QED}

\title[Particle mobility between elastic membranes]
{Particle mobility between two planar elastic membranes: Brownian motion and membrane deformation}

\author{Abdallah Daddi-Moussa-Ider}
\email{abdallah.daddi-moussa-ider@uni-bayreuth.de}
\author{Achim Guckenberger}%

\author{Stephan Gekle}
\affiliation{%
Biofluid Simulation and Modeling, Fachbereich Physik, Universit\"at Bayreuth, \\ Universit\"{a}tsstra{\ss}e 30, Bayreuth 95440, Germany
}%

\date{\today}

\begin{abstract}
We study the motion of a solid particle immersed in a Newtonian fluid and confined between two parallel elastic membranes possessing shear and bending rigidity. 
The hydrodynamic mobility depends on the frequency of the particle motion due to the elastic energy stored in the membrane.
Unlike the single-membrane case, a coupling between shearing and bending exists. 
The commonly used approximation of superposing two single-membrane contributions is found to give reasonable results only for motions in the parallel, but not in the perpendicular direction.
We also compute analytically the membrane deformation resulting from the motion of the particle, showing that the presence of the second membrane reduces deformation.
Using the fluctuation-dissipation theorem we compute the Brownian motion of the particle, finding a long-lasting subdiffusive regime at intermediate time scales.
We finally assess the accuracy of the employed point-particle approximation via boundary-integral simulations for a truly extended particle. They are found to be in excellent agreement with the analytical predictions.
\end{abstract}

\maketitle



\section{Introduction}

The hydrodynamic motion of nanoparticles near elastic membranes plays an essential role in a variety of biological processes and medical applications.
Examples include the potential use of nanoparticles as drug delivery agents \cite{hillaireau09, rosenholm10, Chauhan_2011} or possible adverse health effects due to nanoparticles generated, e.g., from combustion processes and chemical industries \cite{rothen06}.
One of the strongest biological side effects is expected when nanoparticles are taken up by living cells through endocytosis \cite{muhlfeld08, Doherty_2009, Richards_2014, Meinel_2014} for which the hydrodynamically governed approach towards the cell membrane is the essential first step.

Several theoretical and experimental studies have investigated particle dynamics near a single boundary
such as a rigid wall \cite{lorentz07, blake71, cichocki98,  Banerjee_2005, Holmqvist_2006, Choi_2007, Schaffer_2007, CarbajalTinoco_2007, Huang_2007, Kyoung_2008, Sharma_2010, Kazoe_2011, Lele_2011, Dettmer_2014, swan07, jeney08, franosch09, Michailidou_2009, michailidou13, lisicki12, rogers12, lisicki14, Watarai_2014, Yu_2015}
or cylinder \cite{Eral_2010},
a fluid-fluid interface \cite{lee79, berdan81, bickel07, wanggm09, blawz10, Zhang_2013_APL, bickel14, Wang_2014_diffusion},
a partial-slip interface \cite{lauga05, felderhof12}
and an elastic membrane \cite{bickel06, felderhof06,Shlomovitz_2013, Shlomovitz_2014, Boatwright_2014, salez15, Junger_2015, daddi16, saintyves16}. The latter stands apart from both rigid and fluid interfaces as the stretching of the elastic membrane by the moving particle introduces a memory effect in the system.

The influence of a second boundary on particle dynamics has so far been studied only for hard walls. 
The most simple approach is due to Oseen \cite{oseen28} who suggested that the hydrodynamic mobility of a sphere confined between two rigid walls could be approximated by superposition of the leading-order terms from each single wall. 
A more rigorous attempt goes back to Fax\'{e}n  \cite{faxen21} who computed in his dissertation the particle mobility parallel to the walls for the special cases when the particle is in the mid-plane or the quarter-plane between the two hard walls \cite{happel12}.
For an arbitrary location between the two walls, exact solutions for a point particle can be obtained in terms of convergent series using the image technique \cite{liron76, lobry96, bhattacharya02, felderhof06twoMem}.
For a truly extended particle, multipole expansions \cite{swan10} as well as joint analytical-numerical solutions have been presented \cite{ganatos80a, ganatos80b}.
Experimentally, the Brownian dynamics of a spherical particle confined between two parallel rigid walls has been studied using direct imaging measurements in the parallel direction \cite{dufresne01} who found good agreement with Oseen's superposition approximation.
Dynamic-light-scattering \cite{lobry96} and video microscopy combined with optical traps \cite{lin00, Trankle_2016} also found good agreement with theoretical predictions.
Despite the significant progress in this field, the particle motion between two confining elastic interfaces has not been studied so far.
An understanding of how the particle motion is affected by two adjacent elastic walls can be useful to model the diffusion of medical drugs across the extracellular space between neighboring cells \cite{nicholson98} or the transport of macromolecules across endothelial cells that line the surface of blood vessels \cite{maeda13}.

In this paper, we derive an analytical theory for the translational motion of a small solid particle confined between two parallel elastic membranes with both shear and bending resistance.
The theoretical predictions are confirmed by boundary integral simulations.
We find that shearing and bending contributions are intrinsically coupled which is in strong contrast to the single-membrane case where shearing and bending parts are independent and add up linearly to produce the full particle mobility \cite{daddi16}. 
We show that Oseen's often used superposition approximation leads to a reasonably good prediction of the particle mobility only for the parallel, but not for the perpendicular motion, with errors in the mobility correction as high as $55\,\%$.
Furthermore, we investigate the membrane deformation induced by the moving particle and show that the presence of the second membrane significantly reduces deformation compared to the single membrane case.
Finally, the subdiffusive nature of the Brownian motion, which has recently been observed near a single membrane \cite{bickel06, daddi16} is shown to be further enhanced by the presence of the second membrane. 
 
The paper is organized as follows.
In Sec.~\ref{mathematicalFormulation}, we detail the mathematical derivation of the particle mobility for the motion perpendicular and parallel to the membranes.
In Sec.~\ref{simulations}, we present the boundary integral method (BIM) and its implementation together with the procedure that we use to extract the particle mobility.
Particle mobilities, membrane deformations and mean-square displacements are provided in dimensionless form in Sec.~\ref{resultsAndDiscussion}.
Concluding remarks are offered in Sec.~\ref{conclusions}.


\section{Mathematical formulation}
\label{mathematicalFormulation}

\subsection{Problem setup}

\begin{figure}
 \begin{center}
  \includegraphics[scale = 0.4]{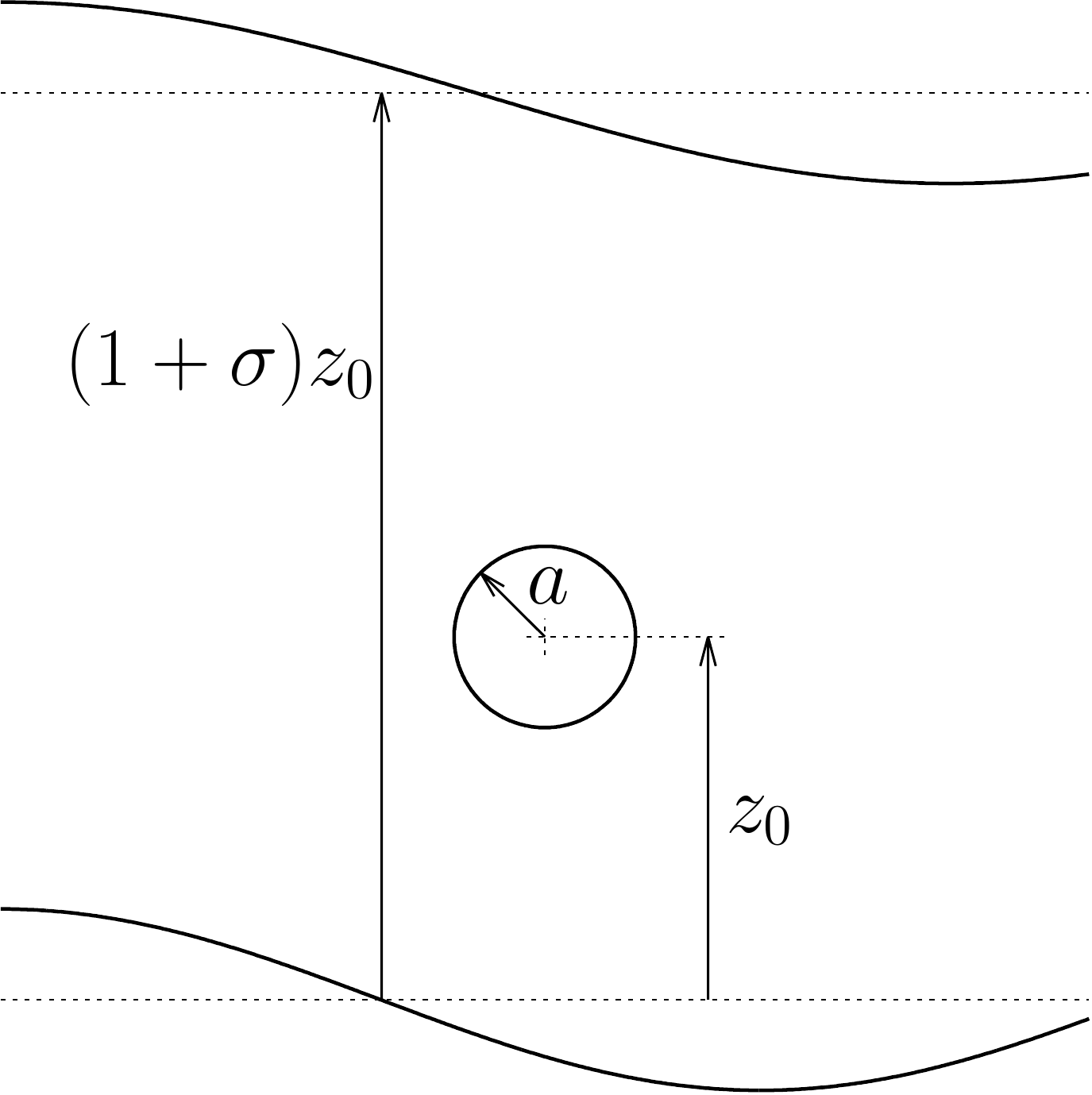}
 \caption{Illustration of the problem setup: A spherical particle of radius $a$ at vertical position $z_0$ moves between two membranes located at $z=0$ and $z=(1+\sigma) z_0$. The membranes have infinite extent in the $x$ and $y$ directions.}
 \label{particleIllustration}
 \end{center}
\end{figure}

We consider a small spherical solid particle of radius~$a$ located at $z=z_0>0$, moving between two parallel elastic membranes having infinite extent in the $xy$ plane.
The first undisplaced  membrane is located at $z=0$ and the second one at $z=(1+\sigma) z_0$, where $\sigma \ge 1$ is a parameter (see Fig.~\ref{particleIllustration} for an illustration.) 
For $\sigma=1$, the particle is at equal distance from the two membranes.
The one-membrane limit may be recovered by taking the limit when $\sigma$ tends to infinity.
Furthermore, the fluid in the whole domain is considered as incompressible and with constant dynamic viscosity $\eta$.

\subsection{Particle mobility}

We aim at computing the particle mobility $\mu_{\alpha \beta}$, a geometry and frequency dependent tensorial quantity that relates the velocity $\vect{V}$ of a solid particle located at $\R_0$ to a force $\vect{F}$ applied on its surface.
Transforming to temporal Fourier space, we have
\begin{equation}
 {V}_\alpha (\omega) = {\mu}_{\alpha \beta} (\R_0, \omega) {F}_\beta (\omega) \, .
\end{equation}
Summation over repeated indices is assumed.
The particle mobility can be split up into two contributions:
\begin{equation}
 \mu_{\alpha \beta}  (\R_0, \omega) = \mu_0 \, \delta_{\alpha \beta} + \Delta \mu_{\alpha \beta} (\R_0, \omega) \, ,
\end{equation}
where $\mu_0 = 1/(6\pi\eta a)$ is the common bulk mobility and $\delta_{\alpha \beta}$ is the Kronecker tensor. The mobility correction $\Delta \mu_{\alpha \beta}$ in the point particle approximation $a \ll z_0$ is expressed as
\begin{equation}
 \Delta \mu_{\alpha \beta} (\R_0, \omega) = \lim_{\R \to \R_0} \left( \G_{\alpha \beta}(\R,\R_0, \omega) - \G_{\alpha \beta}^{(0)} (\R,\R_0) \right) \, ,
 \label{mobilityFromGreensFunction}
\end{equation}
where $\G_{\alpha \beta}$ is the Green's function of the fluid velocity $\vect{v}$ in the presence of the membranes, defined as
\begin{equation}
 v_\alpha (\R, \omega) = \G_{\alpha \beta}(\R,\R_0, \omega) F_\beta (\omega) \, ,
 \label{greenFunctionsDefinition}
\end{equation}
and $\G_{\alpha \beta}^{(0)}$ is the infinite space Green's function, given by
\begin{equation}
 \G_{\alpha \beta}^{(0)} (\R,\R_0) = \frac{1}{8\pi\eta} \left( \frac{\delta_{\alpha \beta}}{s} + \frac{s_\alpha s_\beta}{s^3} \right) \, ,
 \label{infiniteSpaceGreensFunction}
\end{equation}
where $\vect{s} := \R - \R_0$ and $s := |\vect{s}|$.

The particle mobility can be obtained after solving the forced equations of fluid motion for the present boundary conditions.
{We solve them by Fourier-transforming the coordinates parallel to the membranes $x$ and $y$.}
Afterward, the mobility corrections are obtained from Eq.~\eqref{mobilityFromGreensFunction}.
The particle mobility provides the memory kernel of our system and serves as an input for the generalized Langevin equation that governs the diffusional dynamics of the Brownian particle,  as will be described in details in Sec.~\ref{resultsAndDiscussion}.

\subsection{Stokes equations}

For a small Reynolds number,
the fluid velocity $\vect{v}(\R,t)$ and pressure $p(\R,t)$ are governed by the steady Stokes equations 
\begin{align}
 \eta \vect{\nabla}^2 \vect{v} - \vect{\nabla} p +\vect{F} \delta (\R-\R_0) &= 0 \, , \label{eq:Stokes}\\
 \vect{\nabla}\cdot\vect{v} &= 0 \, , \label{eq:Incompress}
\end{align}
where $\vect{F}(t)$ denotes a time-dependent point force (expressed in Newton) acting on the particle position $\R_0 = (0,0,z_0)$. 
Furthermore, $\delta$ signifies the three-dimensional Dirac delta function.
In previous a work \cite{daddi16}, we have shown that the unsteady term in the momentum equation leads to negligible contribution in the mobility correction and is thus not considered here.
The no-slip boundary condition at the membranes provides a direct link between the fluid velocity and the membrane displacement field $\vect{u} (x,y)$, which at leading order in deformation reads 
\begin{equation}
 \left.  \vect{v} = \frac{\Intd  \vect{u}}{\Intd  t}  \right|_{z = 0} \quad \text{and} \quad \left.  \vect{v} = \frac{\Intd  \vect{u}}{\Intd  t}  \right|_{z = (1+\sigma)z_0} \, .
 \label{noSlipCondition}
\end{equation}
Hereafter, we shall denote by $\zm$ the vertical position of each undisplaced membrane, i.e.\ $\zm \in \{0, (1+\sigma)z_0 \}$.
The velocity is continuous at $\zm$ whereas the stretching and bending forces impose a discontinuity in the fluid stress tensor. 
Deformation properties of the RBC membrane are modeled by the Skalak model \cite{skalak73} involving as parameters the shear modulus $\kS$ and the area expansion modulus $\kA$ \cite{daddi16}.
The membrane resists toward bending according to the Helfrich model \cite{helfrich73}.
{Membrane viscosity can in principle be included into our model by adding an imaginary part to the shear modulus $\kS$. 
Yet, since membrane viscosity is a damping term akin to the already included fluid viscosity, we do not expect our results to change significantly if it were to be included.
As we shall see below, the anomalous diffusion on which we focus in the present paper comes from the membrane elasticity providing a memory to the system.} 

With the Skalak and Helfrich models it follows that the linearized tangential and normal fluid stress jumps across the interface are related to the membrane displacement field at $\zm$ by \cite{daddi16}
\begin{subequations}
      \begin{align}
      [\sigma_{z\alpha}] &= -\frac{\kS}{3} \left( \DeltaPara u_\alpha + (1+2C) e_{,\alpha} \right) \,, \quad \alpha \in \{ x,y \} \, , \label{tangentialCondition}\\
      ~[\sigma_{zz}] &= \kB \DeltaPara^2 u_z \, , \label{normalCondition}
      \end{align}
\end{subequations}
where  $[g] = g(z_{\mathrm{m}}^{+}) - g(z_{\mathrm{m}}^{-})$ denotes the jump of a quantity $g$ across the membrane located at $z_{\mathrm{m}}$. 
Furthermore, $C := \kA/\kS$ is the ratio of the area expansion to shear modulus, $\DeltaPara = \partial_{,xx} + \partial_{,yy}$ is the Laplace-Beltrami operator along the membrane and $e=u_{x,x}+u_{y,y}$ is the dilatation.
A comma in indices denotes derivatives.
The components $\sigma_{z\alpha}$ of the stress tensor are expressed by
\begin{equation}
 \sigma_{z\alpha} = -p \, \delta_{z\alpha} + \eta (v_{z,\alpha} + v_{\alpha,z}) \, , \quad \alpha \in \{x,y,z\} \, .
\end{equation}

The Stokes equations can conveniently be solved using a two-dimensional Fourier transform technique \cite{felderhof06, bickel07, daddi16}.
Moreover, the dependence of the membrane shape on the motion history  suggests a temporal Fourier mode analysis.
Here we use the common convention of a negative exponent in the forward Fourier transforms.
As both spacial as well as temporal transformations will be performed, we shall reserve the tilde for the spatially transformed functions while the function and its temporal Fourier transform will be distinguished uniquely by their arguments.

Continuing, it is convenient to adopt the orthogonal coordinate system in which the Fourier transformed vectors are decomposed into longitudinal, transverse and normal components  \cite{bickel07, thiebaud10, daddi16}, denoted by $\vt_l$, $\vt_t$ and $\vt_z$, respectively.
For some given vectorial quantity $\vect{\Qt}$, the passage from the new orthogonal basis to the usual Cartesian basis can be performed via the orthogonal transformation
\begin{equation}
\left( 
	\begin{array}{c}
         \Qt_x \\
         \Qt_y 
        \end{array}
\right)
=
\frac{1}{q}
 \left( 
	\begin{array}{cc}
         q_x & q_y\\
         q_y & -q_x
        \end{array}
\right)
\left( 
	\begin{array}{c}
         \Qt_l \\
         \Qt_t 
        \end{array}
\right) \, ,
\label{transformation}
\end{equation}
where $q_x$ and the $q_y$ are the components of the wavevector $\vect{q}$ and $q:= |\vect{q}|$.
Note that the component $\Qt_z$ along the direction normal to the membranes is left unchanged.

After applying these transformations to Eqs.~\eqref{eq:Stokes} and \eqref{eq:Incompress}, we can eliminate the pressure and obtain two decoupled ordinary differential equations for $\vt_t$ and $\vt_z$, such that
\cite{bickel07, daddi16}
\begin{subequations}
      \begin{align}
       q^2 \vt_t - \vt_{t,zz} &= \frac{F_t}{\eta} \delta (z-z_0) \, , \label{transverseEquation}\\
       \vt_{z,zzzz} - 2q^2 \vt_{z,zz} + q^4 \vt_z &= \frac{q^2 F_z}{\eta} \delta(z-z_0) \nonumber \\
		     \hphantom{\vt_{z,zzzz} - 2q^2 \vt_{z,zz} + q^4 \vt_z} & \hphantom{{}={}} + \frac{iq F_l}{\eta} \delta' (z-z_0) \, , \label{normalEquation}
      \end{align} \label{systemOfEquations}
\end{subequations}
where $\delta'$ stands for the derivative of the Dirac delta function.
The incompressibility equation~\eqref{eq:Incompress} allows for the determination of $\vt_l$ from  $\vt_z$ such that
\begin{equation}
 \vt_l = \frac{i \vt_{z,z}}{q} \, . \label{longitudinalFromNormal}
\end{equation}

For the sake of amenable mathematical equations, we will only consider the case that the two membranes have the same elastic and bending properties. 
Indeed, this is usually encountered in blood vessels where the RBCs posses similar physical properties.
After some algebra it can be shown that the stress jump due to shear and area expansion from Eq.~\eqref{tangentialCondition} imposes the following discontinuities at $\zm$ \cite{daddi16}:
\begin{subequations}
 \begin{align}
   [\vt_{t,z}] &= \left. -iB\alpha  q^2 \vt_t \right|_{z = \zm} \, , \label{transverseVeloPrime}\\
   ~[\vt_{z,zz}] &= \left. -{4i\alpha q^2} \vt_{z,z} \right|_{z = \zm} \, ,\label{normalVeloSecond}
 \end{align}
\end{subequations}
where $\alpha := \kS/(3 B\eta\omega)$ with $B := 2/(1+C)$ is a characteristic length for shear and area expansion. 
The normal stress jump given by Eq.~\eqref{normalCondition} leads to
\begin{equation}
 [\vt_{z,zzz}] = \left. 4i \alphaB^3 q^6 \vt_z \right|_{z = \zm} \, ,\label{normalVeloThird}
\end{equation}
where $\alphaB := (\kB/(4\eta\omega))^{1/3}$ is a characteristic length for bending.

\subsection{Solutions}

The basic approach for solving such a system of equations \eqref{systemOfEquations} and \eqref{longitudinalFromNormal} to obtain the particle mobility was detailed in an earlier work \citep{daddi16}. Here we only outline the major differences and steps.

Since the system is isotropic with respect to the $x$ and $y$ directions the mobility tensor only contains diagonal components. 
The normal-normal component $\Gt_{zz}$ can be obtained from solving Eq.~\eqref{normalEquation} in which only the normal force $F_z$ is considered, i.e.\ $F_l=0$.
By applying the appropriate boundary conditions at $\zm$ and $z_0$, the integration constants are readily determined.
At $z=\zm$, the normal velocity $\vt_z$ and its first derivative are continuous whereas the second and third derivatives are discontinuous because of shearing and bending, as prescribed in Eqs.~\eqref{normalVeloSecond} and \eqref{normalVeloThird} respectively.
At the point force position, i.e.\ at $z=z_0$, the normal velocity and its first and second derivatives are continuous while the Dirac delta function imposes the discontinuity of the third derivative (see Eq.~\eqref{normalEquation}).

For the motion parallel to the membranes, it is sufficient to consider a force $F_x$ and solve for the Green's function component $\Gt_{xx}$.
The latter can be expressed by employing Eq.~\eqref{transformation} via
\begin{equation}
  {\Gt}_{xx} (q,\phi,\omega) =   {\Gt}_{tt} (q,\omega) \sin^2 \phi  + {\Gt}_{ll} (q,\omega) \cos^2 \phi \, ,
  \label{G_xx_qPhiOmega}
\end{equation}
where $\phi := \arctan (q_y/q_x)$.
Accordingly, the determination of $\Gt_{xx}$ requires two steps.
First, the transverse-transverse component $\Gt_{tt}$ is determined from solving Eq.~\eqref{transverseEquation}.
The transverse velocity $\vt_t$ is continuous at the membranes whereas shearing imposes the discontinuity of the first derivative as prescribed by Eq.~\eqref{transverseVeloPrime}.
At $z=z_0$, the transverse velocity is continuous  while its first derivative is discontinuous because of the Dirac delta function (see Eq.~\eqref{transverseEquation}).
Second, the  normal velocity component $\vt_z$ is determined as an intermediate step from solving first Eq.~\eqref{normalEquation} by only considering the longitudinal force $F_l$, i.e.\ $F_z = 0$.
In this situation, the Dirac delta function imposes the discontinuity of the second derivative at $z_0$ whereas the third derivative is continuous.
Afterward, the velocity component $\vt_l$ is immediately recovered thanks to the incompressibility equation~\eqref{longitudinalFromNormal}, giving access to the longitudinal-longitudinal component~$\Gt_{ll}$.

What remains for the determination of the particle mobility is to apply the spatial inverse Fourier transform by integrating over $\phi$ and the wavenumber $q$.
In the point particle approximation, the mobility correction can readily be calculated by subtracting the bulk term and taking the limit when $\R$ tends to $\R_0$, as described by Eq.~\eqref{mobilityFromGreensFunction}.


{For convenience, we define the subscripts $\perp$ and $\parallel$ to denote the tensorial components $zz$ and $xx$, respectively.
The $yy$ component of the mobility tensor is identical to the $xx$ component.}
Moreover, we define $k_\perp^{\sigma} (\beta, \betaB)$ and $k_\parallel^{\sigma} (\beta, \betaB)$, two frequency dependent complex quantities 
which are related to the first order correction in the mobility via
\begin{equation}
 \frac{\Delta \mu_\alpha (z_0, \omega)}{\mu_0} = -k_\alpha^{\sigma} (\beta, \betaB) \frac{a}{z_0} \, ,  \quad \alpha \in \{ \perp, \parallel \} \, ,
 \label{DeltaMuDefinition}
\end{equation}
where  $\beta := 2z_0/\alpha \sim \omega$ and $\betaB := 2z_0/\alphaB \sim \omega^{1/3} $ are two dimensionless frequencies related to the shear and bending effects, respectively.
Analytical expressions for $k_\alpha^{\sigma} (\beta, \betaB)$ can be obtained with computer algebra software, but they are not listed here due to their complexity and lengthiness. 
\footnote{See Supplemental Material at [URL will be inserted by publisher] for a Maple script (Maple 17 or later) providing the particle mobility corrections in both directions of motion.}
{These expressions are the basis for the computation of the Brownian motion and therefore constitute one of the central results of our work.}

{We proceed to investigate the limiting case of Eq.~\eqref{DeltaMuDefinition} in which both shearing and bending modulus tend to infinity and therefore $\beta$ and $\betaB$ both tend to zero.
In this case, which physically represents a hard wall, the general expression for $k_{\alpha}^{\sigma}$ as it appears in Eq.~\eqref{DeltaMuDefinition} reduces to}
\begin{widetext}
\begin{subequations}
 \begin{align} 
  k_{\perp}^{\sigma} (0,0) &= \int_{0}^{\infty} \frac{3}{4\Gamma} \left( \phi_{+}^{1}e^{2\sigma u}-\phi_{-}^{1}e^{-2\sigma u} + \phi_{+}^{\sigma} e^{2u}-\phi_{-}^{\sigma} e^{-2u} +e^{-2(1+\sigma)u} -\psi_{+} \right) \Intd u \, , \\
  k_{\parallel}^{\sigma} (0,0)  &= \int_{0}^{\infty} \left( \frac{3}{8\Gamma} \left( \phi_{-}^{1}e^{2\sigma u}-\phi_{+}^{1}e^{-2\sigma u} + \phi_{-}^{\sigma} e^{2u}-\phi_{+}^{\sigma} e^{-2u} +e^{-2(1+\sigma)u} -\psi_{-} \right) -\frac{3}{4} \frac{e^{2u} + e^{2\sigma u} - 2}{e^{2(1+\sigma)u} - 1} \right) \Intd u \, ,
 \end{align}
 \label{muHardWall}
 \end{subequations}
 \end{widetext}
where we defined
\begin{subequations}
\begin{align}
 \phi_{\pm}^{\sigma} &:= \sigma u (\sigma u \pm 1) + \frac{1}{2} \, ,\\
 \psi_{\pm} &:= 1 + 2(1+\sigma)^2 u^2 \pm 2(1+\sigma) (1+2\sigma u^2) u \, ,\\
 \Gamma &:= 1+2(1+\sigma)^2 u^2 -\cosh \left( 2(1+\sigma) u \right) \, .
\end{align}
 \end{subequations}
 {
Expressions \eqref{muHardWall} are valid for arbitrary positions of the upper wall given by $(1+\sigma)z_0$.
For specific values of $\sigma$ we recover three results obtained earlier:}
First, the single hard wall limits $k_\perp^{\infty}(0,0) = 9/8$ and $k_\parallel^{\infty}(0,0) = 9/16$  \cite{lorentz07, lee79} are obtained for $\sigma\to\infty$.
Second, the two wall case for $\sigma = 1$ and $\sigma=3$ lead to the first order correction terms for the parallel motion as computed by Fax\'{e}n \cite{happel12}, namely $k_\parallel^{1}(0,0) \approx 1.0041$ and $k_\parallel^{3}(0,0) \approx 0.6526$. 
Third, we find the result by Felderhof \cite{felderhof06twoMem} for the perpendicular motion, $k_\perp^{1}(0,0) \approx 1.4516$.

\subsection{Coupling of shear and bending contributions}
\label{sec:coupling}

In this subsection we address one particular aspect of the boundary conditions for the two membranes.
In our recent work \cite{daddi16} we found that the particle mobility near a single elastic membrane could be expressed as the linear combination of the two independent shear and bending contributions.
For the two membrane case as discussed in the present work, however, the solution of Eq.~\eqref{normalEquation} requires to simultaneously consider the boundary conditions stated by Eqs.~\eqref{normalVeloSecond} and \eqref{normalVeloThird}.
This is a qualitative difference compared to the one membrane case.

To see this, consider two different setups, one with only bending resistance ($\alpha=0$) and one with only shear resistance ($\alphaB=0$). Furthermore, let the corresponding perpendicular velocities be denoted by $\vt_z^{\mathrm{B}}$ and $\vt_z^{\mathrm{S}}$, respectively. 
If the expression $\vt_z^{\mathrm{S}} + \vt_z^{\mathrm{B}} - \vt_z^{\mathrm{bulk}}$ should be the solution of two membranes with shear and bending resistance, 
it would have to fulfill the boundary conditions~\eqref{normalVeloSecond} and~\eqref{normalVeloThird}. 
This is true if and only if
\begin{equation}
  \left. \vt_{z,z}^{\mathrm{B}} \right|_{z = \zm} = 0 \quad \text{and} \quad   \left. \vt_{z}^{\mathrm{S}} \right|_{z = \zm} = 0 \,,
 \label{conditionSplitUpSolutions}
\end{equation}
which is in general satisfied only in the one membrane limit. {
As a result, the contributions from shearing and bending cannot be added independently on top of each other in the resulting mobility corrections, defined by Eq.~\eqref{DeltaMuDefinition}.}



\subsection{Computation of membrane deformations}

A force acting on a particle will induce a motion in the fluid.
As a result, the imbalance in the stress tensor across the membranes leads to their deformation.
In this subsection we compute the deformation resulting from a time dependent point force located at $z_0$, whereas the force is oriented perpendicularly or parallel to the membranes.
Once the fluid velocity field is computed in the whole domain, the displacement field for each membrane  can be obtained via Eq.~\eqref{noSlipCondition}. For each membrane we define a frequency and wavevector dependent reaction tensor $\psit_{\alpha \beta}$ as
\begin{equation}
\ut_\alpha (\vect{q},\omega) = \psit_{\alpha \beta} (\vect{q}, \omega) F_{\beta} (\omega) \, . 
\label{reactionTensorDefinition}
\end{equation}

For the perpendicular motion, the radial symmetry suggests that the displacement vector will have a normal component $u_z$ and a radial component $u_r$.
By performing the spatial inverse Fourier transform for a radially symmetric function \cite{baddour09}, we immediately get the normal-normal component of the reaction tensor in real-space:
\begin{equation}
 \psi_{zz} (\rho, \omega) = \frac{1}{2\pi} \int_{0}^{\infty} \psit_{zz} (q,\omega) \, J_0 (\rho q) \, q \, \Intd q \, ,
 \label{psi_zz}
\end{equation}
where $\rho := \sqrt{x^2+y^2}$ and $J_0$ is the zeroth-order Bessel function.

To compute the radial-normal component, we first note that from the transformation equations~\eqref{transformation} $\psit_{xz} = \psit_{lz} \cos \phi$ since $\psit_{tz} = 0$ in virtue of the decoupled nature of Eqs.~\eqref{transverseEquation} and~\eqref{normalEquation}.
Thus, the spatial inverse Fourier transform applied to the non-radially symmetric function $\psit_{xz} (q,\phi,\omega)$ leads to
\begin{equation}
 \psi_{rz} (\rho, \omega) = \frac{i}{2\pi}  \int_{0}^{\infty} \, \psit_{lz} (q,\omega) \, J_1 (\rho q) \, q \, \Intd q \, ,
 \label{psi_rz}
\end{equation}
using the fact that  $\psi_{xz} = \psi_{rz} \cos \theta$ and $\psi_{yz} = \psi_{rz} \sin \theta$ where $\theta := \arctan (y/x)$.

Let us consider next the deformation due to a time dependent point force parallel to the membranes.
Due to the symmetry it suffices to consider a force applied along the $x$-direction.
Furthermore, this force can be decomposed into a longitudinal component $F_l = F_x \cos \phi$ and a transverse component $F_t = F_x \sin \phi$.
For the normal-tangential component $\psi_{zx}$, it follows from the transformation equations~\eqref{transformation} that $\psit_{zx} = \psit_{zl} \cos \phi$ since $\psit_{zt} = 0$ for the same reason as $\psit_{tz}$.
Therefore, the inverse Fourier transform back into real space gives
\begin{equation}
 \psi_{zx} (\rho, \theta, \omega) = \frac{i \cos \theta}{2\pi}   \int_{0}^{\infty} \, \psit_{zl} (q,\omega) \, J_1 (\rho q) \, q \, \Intd q \, ,
 \label{psi_zx}
\end{equation}
meaning that the vertical deformation is maximal in the plane $y=0$ containing the support of the vector force, and vanishes in the plane $x=0$ perpendicular to it.

To compute the lateral stretching of the membrane due to a parallel force on the particle, we require the components $\psi_{xx}$ and $\psi_{yx}$ giving access to the two in-plane displacements $u_x$ and $u_y$, respectively.
It follows immediately from applying the transformation equations~\eqref{transformation} together with the definition of the reaction tensor Eq.~\eqref{reactionTensorDefinition} that
\begin{equation}
\psit_{xx} (q,\phi,\omega) =   \psit_{ll} (q,\omega) \cos^2 \phi +  \psit_{tt} (q, \omega)  \sin^2 \phi \, ,
\end{equation}
leading after spatial inverse Fourier transform to
\begin{equation}
\begin{split}
 \psi_{xx} (\rho, \theta, \omega) &= \frac{1}{4\pi}  \int_0^{\infty} \bigg( \left(\psit_{ll}(q,\omega)+\psit_{tt}(q,\omega) \right) J_0 (\rho q) \\
 &+ \left(\psit_{tt}(q,\omega)-\psit_{ll} (q,\omega) \right) J_2 (\rho q) \cos 2\theta \bigg) \, q \, \Intd q \, .
 \end{split}
 \label{psi_xx}
\end{equation}
Similar, for $\psi_{yx}$ we have
\begin{equation}
 \psit_{yx} (q,\phi,\omega) =  \left( \psit_{ll} (q,\omega) - \psit_{tt} (q,\omega) \right) \cos \phi \sin \phi \, , 
\end{equation}
whose inverse Fourier transform is
\begin{equation}
	\begin{split}
		 \psi_{yx} & (\rho, \theta, \omega)	 = \frac{\sin 2\theta }{4\pi} \\
		  & \times \int_0^{\infty}  \left(\psit_{tt}(q,\omega)-\psit_{ll}(q,\omega) \right) J_2 (\rho q) \, q \, \Intd q    \, .
	\end{split}
 \label{psi_yx}
\end{equation}

Although not transparent from Eq.~\eqref{psi_xx}, the deformation in the $x$-direction is maximal in the plane $y=0$ and minimal in the plane $x=0$.
On the other hand, deformation is maximal for the $y$-direction in the bisector planes $y= \pm x$, and vanishes in the planes $x=0$ and $y=0$.
Under the action of an arbitrary time dependent point force $\vect{F}(t)$, the membrane deformation can subsequently be obtained by applying the temporal inverse Fourier transform.


\section{Simulations}
\label{simulations}

\subsection{Boundary Integral Method}

For the simulations we use the boundary integral method (BIM) \cite{PozrikidisBook92} whose foundation is the steady Stokes equations.
The core idea is to write them as an integral equation, made possible by the fact that we deal with a linear equation.
However, treating rigid objects in the direct formulation is difficult and inefficient since it would lead to a Fredholm equation of the first kind.
Instead, we employ an extension called the completed double layer boundary integral equation method (CDLBIEM) \cite{kohr04,Zhao2012a}.
For the system with the two membranes the equations read
\begin{subequations} \label{eq:CDL}
	\begin{align}
		v_j(\vect{x}) &= H_j(\vect{x}) \, , \quad \vect{x} \in S_{\mathrm{m}} \, , \\
		\frac{1}{2} \phi_j(\vect{x}) + \sum_{i=1}^{6} \varphi_j^{(i)}(\vect{x}) \braket{\vect{\varphi}^{(i)}, \vect{\phi}} &= H_j(\vect{x}) \, , \quad \vect{x} \in S_\mathrm{p} \, .
	\end{align}
\end{subequations}
Here, $S_{\mathrm{m}} := S_{\mathrm{m}_1} \cup S_{\mathrm{m}_2}$ where $S_\mathrm{m1}$ and $S_\mathrm{m2}$ are the surfaces of the two elastic membranes, and $S_\mathrm{p}$ is the surface of the rigid particle of radius $a$.
{The two membranes have a square shape with a length of $300a$.}
$\vect{v}$ represents the velocity on the membranes while $\vect{\phi}$ denotes the so-called double layer density function on $S_\mathrm{p}$.
The latter is an unphysical auxiliary field. However, the corresponding physical velocity can be retrieved via 
\begin{equation}\label{eq:CDL:Vel}
	V_j(\vect{x}) = \sum_{i=1}^{6} \varphi_j^{(i)}(\vect{x}) \braket{\vect{\varphi}^{(i)}, \vect{\phi}} \, , \quad \vect{x} \in S_\mathrm{p} \, .
\end{equation}
where the $\vect{\varphi}^{(i)}$ are known functions representing the six possible rigid body movements of the solid particle \cite{kohr04}. 
The brackets denote the inner product in the vector space of real functions whose domain is $S_\mathrm{p}$.
Continuing, the function $H_j$ with $j=1,2,3$ is given by
\begin{equation}
	H_j(\vect{x}) := -  (N_{\mathrm{m}} \Delta \vect{f})_j (\vect{x}) -  (K_\mathrm{p} \vect{\phi})_j(\vect{x}) + \mathcal{G}_{jk}^{(0)}(\vect{x}, \vect{x}_\mathrm{c}) F_k   \, ,
	\label{functionH_BIM}
\end{equation}
with $\vect{x}_\mathrm{c}$ being the particle centroid.
We defined the single layer integral via
\begin{equation}
	(N_{\mathrm{m}} \Delta \vect{f})_j (\vect{x}) := \int_{S_{\mathrm{m}}} \Delta f_i(\vect{y}) \mathcal{G}_{ij}^{(0)}(\vect{y}, \vect{x}) \, \Intd S(\vect{y})
\end{equation}
{where integration over both membrane surfaces $S_{\mathrm{m}} := S_{\mathrm{m}_1} \cup S_{\mathrm{m}_2}$ needs to be performed.} The double layer integral is
\begin{equation}
	(K_\mathrm{p} \vect{\phi})_j(\vect{x}) := \oint_{S_\mathrm{p}} \phi_i(\vect{y}) \mathcal{T}_{ijk}^{(0)}(\vect{y},\vect{x}) n_k(\vect{y}) \, \Intd S(\vect{y}) \, .
\end{equation}
The remaining quantities are the jump of the traction $\Delta \vect{f}$ across the membranes, the known force $\vect{F}$ acting on the rigid particle, the outer normal vector $\vect{n}$, the free-space Stokeslet as defined in Eq.~\eqref{infiniteSpaceGreensFunction}, and the corresponding Stresslet
\begin{equation}
 \mathcal{T}_{ijk}^{(0)}(\vect{y},\vect{x}) := -\frac{3}{4\pi} \frac{s_i s_j s_k}{s^5} \, ,
\end{equation}
with $\vect{s} := \vect{y} - \vect{x}$ and $s := |\vect{s}|$.

Given the traction jump $\Delta \vect{f}$ (computed from the current deformation as explained in the appendix) and the force $\vect{F}$ as input, equations~\eqref{eq:CDL} constitute a set of Fredholm integral equations of the second kind for the unknown velocity $\vect{v}$ on the membranes and the density $\vect{\phi}$ on the rigid particle.
To solve this equation numerically, we discretize all surfaces with flat triangles.
For the rigid particle, this is done by consecutively refining an icosahedron \cite{Kruger2011} while gmsh \cite{geuzaine09} was used for the membranes: The quadratic planes were meshed with triangles, with increasing resolution towards their center.
We perform the integration numerically by a Gaussian quadrature with seven points per triangle \cite{Cowper1973} together with linear interpolation of nodal values across each triangle \cite{PozrikidisBook92}.
The singularities appearing in the single layer integral are treated via the polar integration rule \cite{Pozrikidis1995}, while the singularities of the double layer integral are eliminated by the standard singularity subtraction scheme \cite{PozrikidisBook92}. 
With this the integral equation can be evaluated at all nodes, forming a dense and asymmetric linear system of equations which is then subsequently solved by GMRES \cite{Saad1986}.
The residuum of the solver was fixed to $10^{-4}$.
This provides us with the velocity $\vect{v}$ at each node of the two membranes and, after application of equation~\eqref{eq:CDL:Vel}, also of the rigid particle.
The dynamical evolution of the system is hence obtained by solving the kinematic condition \cite{Pozrikidis2001a}
\begin{equation}
	\frac{\Intd  \vect{x}}{\Intd  t} = \vect{v}(\vect{x})
\end{equation}
with the explicit Euler scheme.
We chose a step size that is dependent on the wiggling frequency of the force (cf.\ the next section).


\subsection{Obtaining the mobility from BIM simulations}

In order to obtain the frequency dependent particle mobility from the BIM simulations, an oscillating force $\vect{F}(t)=\vect{A} e^{i\omega_0 t}$ of amplitude $\vect{A}$ and frequency $\omega_0$ is exerted on the particle, in the direction  perpendicular or parallel to the membranes.
After an initial transitory evolution, the particle begins to oscillate with the same frequency as $\vect{V} e^{i(\omega_0 t + \delta)}$. The velocity amplitude $\vect{V}$ and the phase shift $\delta$ can be accurately obtained by fitting the numerically recorded velocity.
For that, we use a nonlinear least-squares solver based on the trust region method \cite{conn00}.
The complex frequency dependent particle mobility can then be evaluated from
\begin{equation}
 \mu_{\alpha} (\omega_0) = \frac{V_{\alpha}}{A_{\alpha}} e^{i\delta} \, .
\end{equation}

For each applied frequency, the force is exerted during three periods in order to ensure that the steady state has been reached properly.
Therefore, lower frequencies require larger computation times.
For instance, for $\beta=10^{-3}$, which is the lowest scaled frequency that we use in our simulations, each period requires around 30 hours using 40 CPUs. 


\section{Results and discussion}
\label{resultsAndDiscussion}

\subsection{Particle mobility}

\begin{figure}
 \begin{center}
  \includegraphics[scale = 0.45]{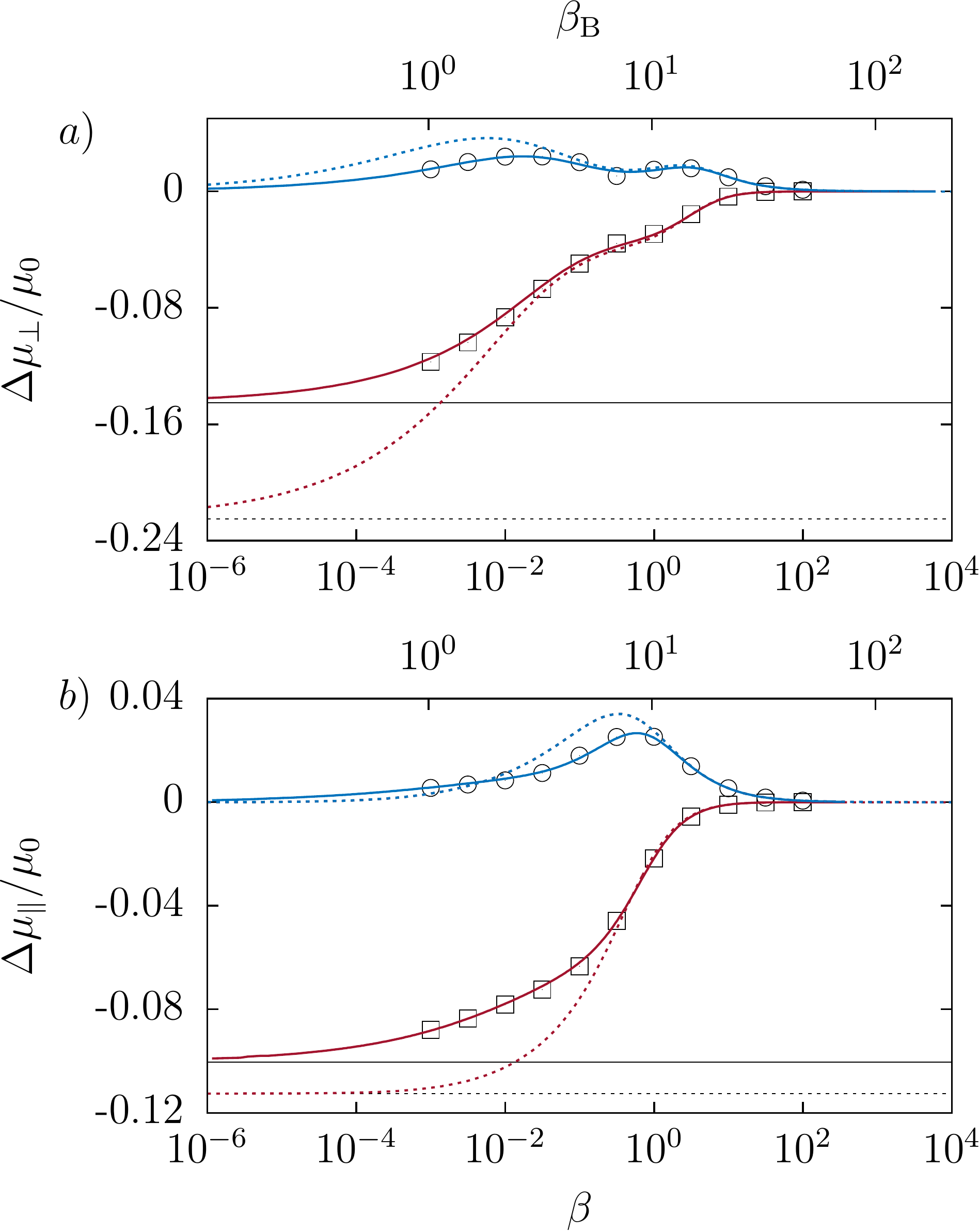}
 \caption{(Color online) The scaled frequency dependent correction to the particle mobility versus the dimensionless frequencies $\beta$ (lower axis) and $\betaB$ (upper axis) for the perpendicular $(a)$ and parallel $(b)$ motions. 
 Here, the particle is equidistant from both membranes ($\sigma = 1$) and located at $z_0 = 10a$. 
 The theoretical predictions {from Eq.~\eqref{DeltaMuDefinition}} are shown as red lines (real part) and blue lines (imaginary part) whereas the BIM simulation results are marked as rectangles (real part) and circles (imaginary part).
 Dashed lines represent the superposition approximation by summing up the contributions of each membrane independently as given by Eq.~\eqref{oseenElastic}.
 The solid horizontal lines indicate the two-hard-wall limits ($-1.4516 a/z_0$ and $-1.0041 a/z_0$ for the perpendicular and parallel motions, respectively) and the dotted horizontal lines result from the superposition approximation of the hard wall as stated in Eq.~\eqref{oseensApproximation}.
 For the other simulation parameters, see main text.}
 \label{CDL_twoMem}
 \end{center}
\end{figure}

We consider a spherical particle equally distant from both membranes $(\sigma=1)$ and located at $z_0 = 10a$.
The membrane reduced bending modulus, defined as $E_\mathrm{B} := \kB / (a^2 \kS)$, is taken to be $E_\mathrm{B} = 1/2$.
We examine the case where $C=1$, for which the Skalak model is equivalent to the common neo-Hookean model \cite{ramanujan98} for small deformations \cite{lac04}.
As shown in Fig.~\ref{CDL_twoMem},
the analytical and numerical results are in very good agreement for the whole range of the applied frequencies, similar as in earlier work for a single membrane \cite{daddi16}.

For a frequency of zero, the imaginary part vanishes. On the other hand, the real part reaches its minimal value which corresponds to the two-hard-walls limit, namely $-1.4516 a/z_0$ and $-1.0041 a/z_0$ for the perpendicular and parallel motions, respectively. This is in agreement with earlier works \cite{felderhof06twoMem, happel12}.

By taking the frequency to infinity, both the real and imaginary parts of the particle mobility correction vanish and one recovers the bulk behavior in which the particle motion is no longer affected by the presence of the membranes. In between, the imaginary part peaks around $\beta \approx 1$ and $\betaB \approx 1$ for the perpendicular motion, and around $\beta \approx 1$ for the parallel motion.
The peak around $\beta \approx 1$, which is observed in both directions, is a shearing signature in the mobility correction, 
whereas the frequency peak around $\betaB \approx 1$ is a signature of bending.
The latter is found to be insignificant in the parallel motion.
Physically, the peak frequencies correspond to the situation where the particle-membranes system naturally vibrates to absorb more energy.

As already remarked, a commonly used approximation to compute mobilities between two walls is Oseen's approach \cite{oseen28} which assumes that the mobility corrections can be approximated by superposing the contributions from each membrane independently as
\begin{equation}
 \frac{\Delta \mu_{\alpha} (z_0, \omega)}{\mu_0} = - \left( k_{\alpha}^{\infty} (\beta,\betaB) + \frac{k_{\alpha}^{\infty} (\sigma \beta, \sigma \betaB)}{\sigma} \right) \frac{a}{z_0} \, ,
 \label{oseenElastic}
\end{equation}
which reduces in the two-hard-wall limit to
\begin{equation}
 \frac{\Delta \mu_{\alpha} (z_0, 0)}{\mu_0} = -k_{\alpha}^{\infty} (0,0) \left( 1 + \frac{1}{\sigma}\right) \frac{a}{z_0} \, .
 \label{oseensApproximation}
\end{equation}

The superposition approximation as given by Eq.~\eqref{oseenElastic} {for the elastic membranes} is compared in Fig.~\ref{CDL_twoMem} against our analytical predictions {from Eq.~\eqref{DeltaMuDefinition} (see also the Supporting Material)} and numerical simulations in order to assess its accuracy.
For the perpendicular motion, we observe that it only agrees well with the analytical predictions and the BIM simulations for frequencies $\betaB > 1$.
At lower frequencies substantial disagreement is observed which, in the limit of a vanishing frequency (hard-walls), amounts to $55\,\%$. 
{This deviation is due to the fact that the superposition approximation allows the fluid to drain away, as the no-slip boundary condition is no longer satisfied at both membranes simultaneously.
As expected, it is therefore more pronounced the more the membrane deforms, i.e.\ for smaller frequencies.}
On the other hand, for the motion parallel to the membranes, the agreement is reasonable down to a dimensionless frequency $\beta$ of order unity.
Below that, however, a significant mismatch between the two curves is observed. In the limit for a vanishing frequency, a relative deviation of $12\,\%$ from Fax\'{e}n's value is obtained.
All in all, the superposition approximation consistently underestimates the particle mobility.


\begin{figure*}
 \begin{center}
  \includegraphics[scale = 0.9]{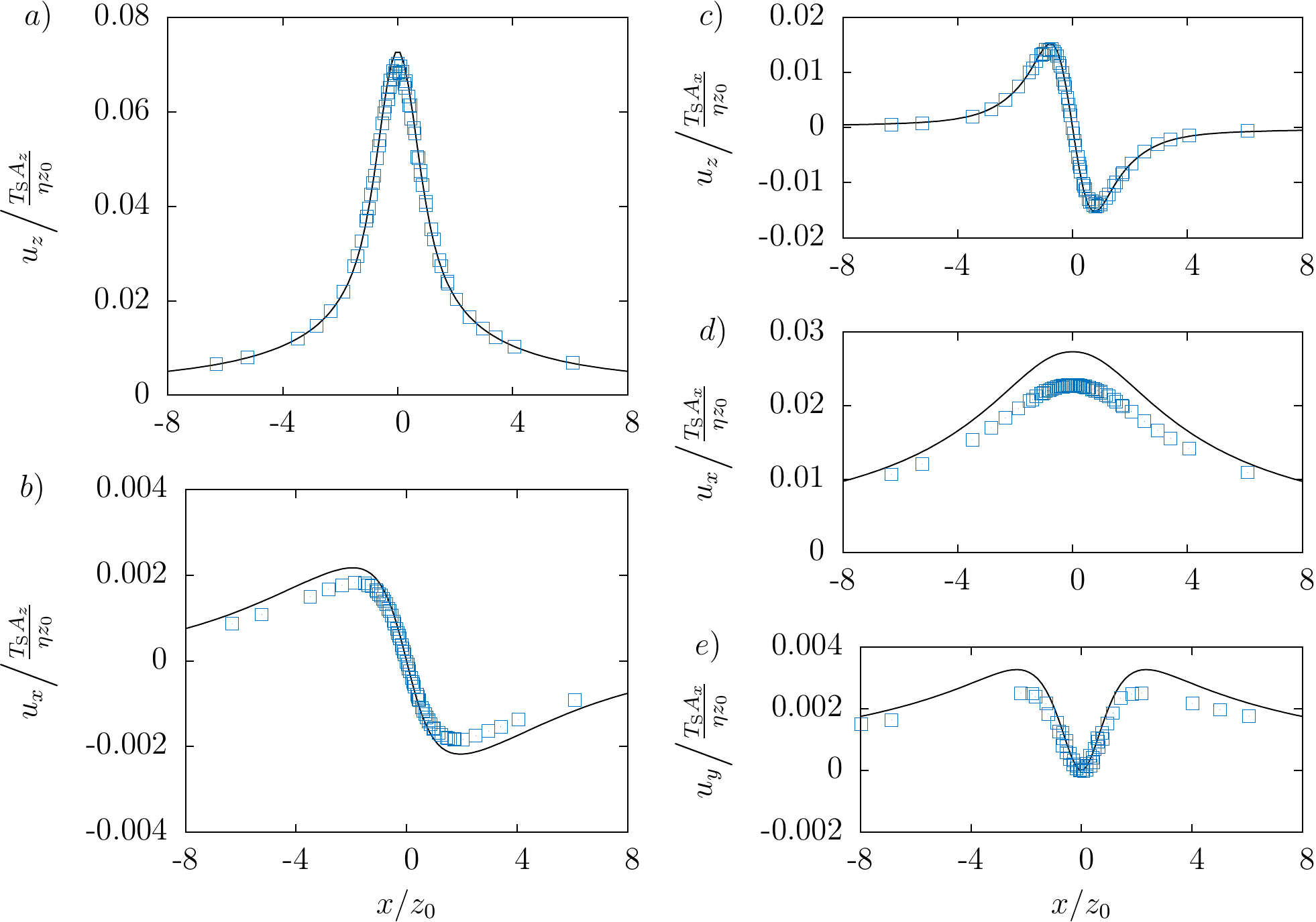}
 \caption{(Color online) Comparison between analytical predictions (solid lines) and numerical simulations (symbols) of the scaled membrane displacement as given by Eq.~\eqref{displacementExpoForce} for the motion perpendicular $(a$ and $b)$ and parallel  $(c, d$ and $e)$ to the membranes, for the parameters given in Fig.~\ref{CDL_twoMem} $(\sigma=1)$. 
 In this example,  we take $\omega_0 \TS = 1$ and $t\omega_0 = \pi/2$.  }
 \label{shapeModulusAll_CDL}
 \end{center}
\end{figure*}


\subsection{Membrane deformation}

\begin{figure}
 \begin{center}
  \includegraphics[scale = 0.4]{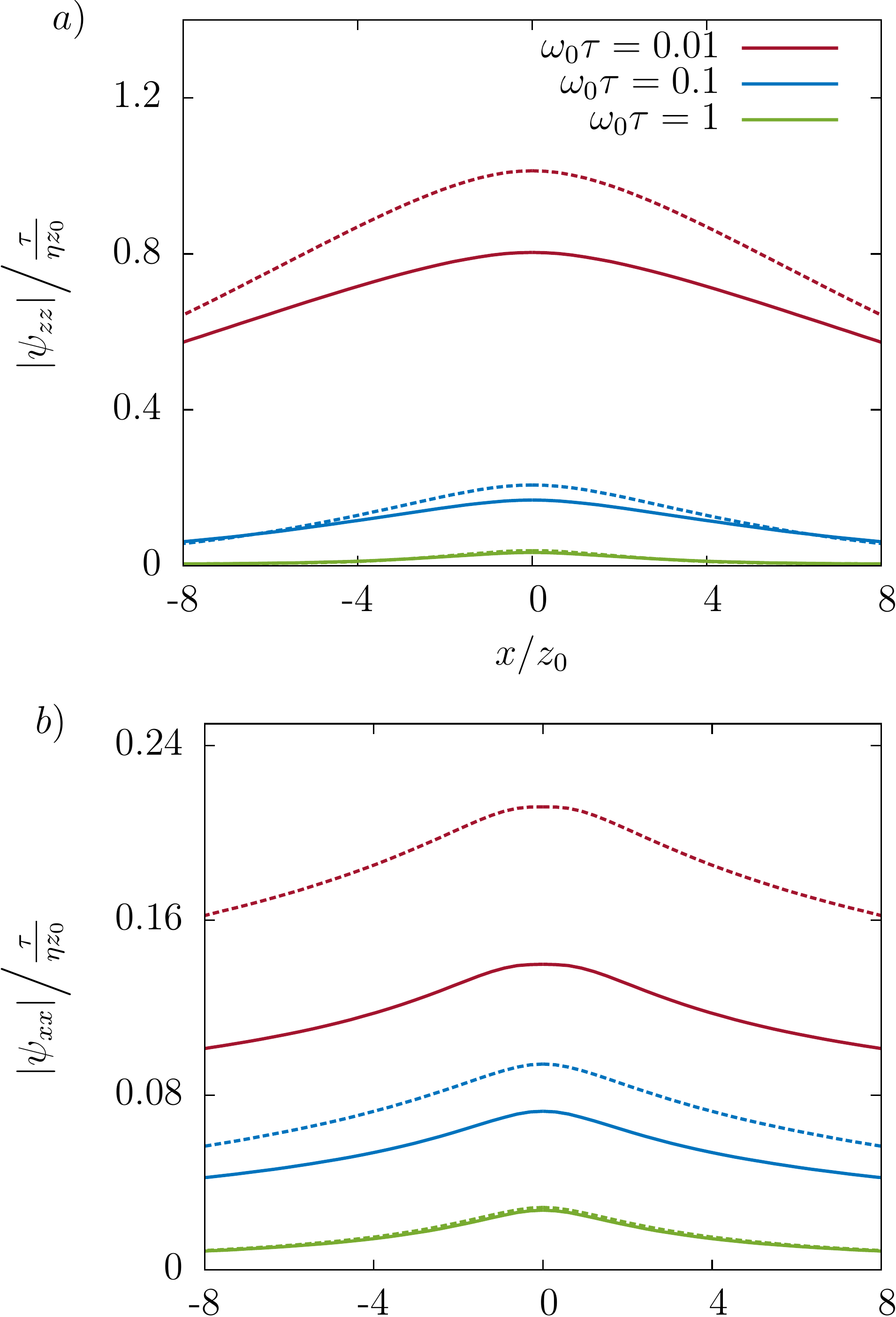}
 \caption{(Color online) Effect of the oscillation frequency on the amplitude of the reaction tensor's $z$-$z$ component $(a)$ and $x$-$x$ component $(b)$ for $\sigma = 1$ (solid line) and $\sigma=\infty$ (dashed line) with $z_0 = (3\kB/(2\kS))^{1/2}$.}
 \label{omega_Effect}
 \end{center}
\end{figure}

\begin{figure}
 \begin{center}
  \includegraphics[scale = 0.4]{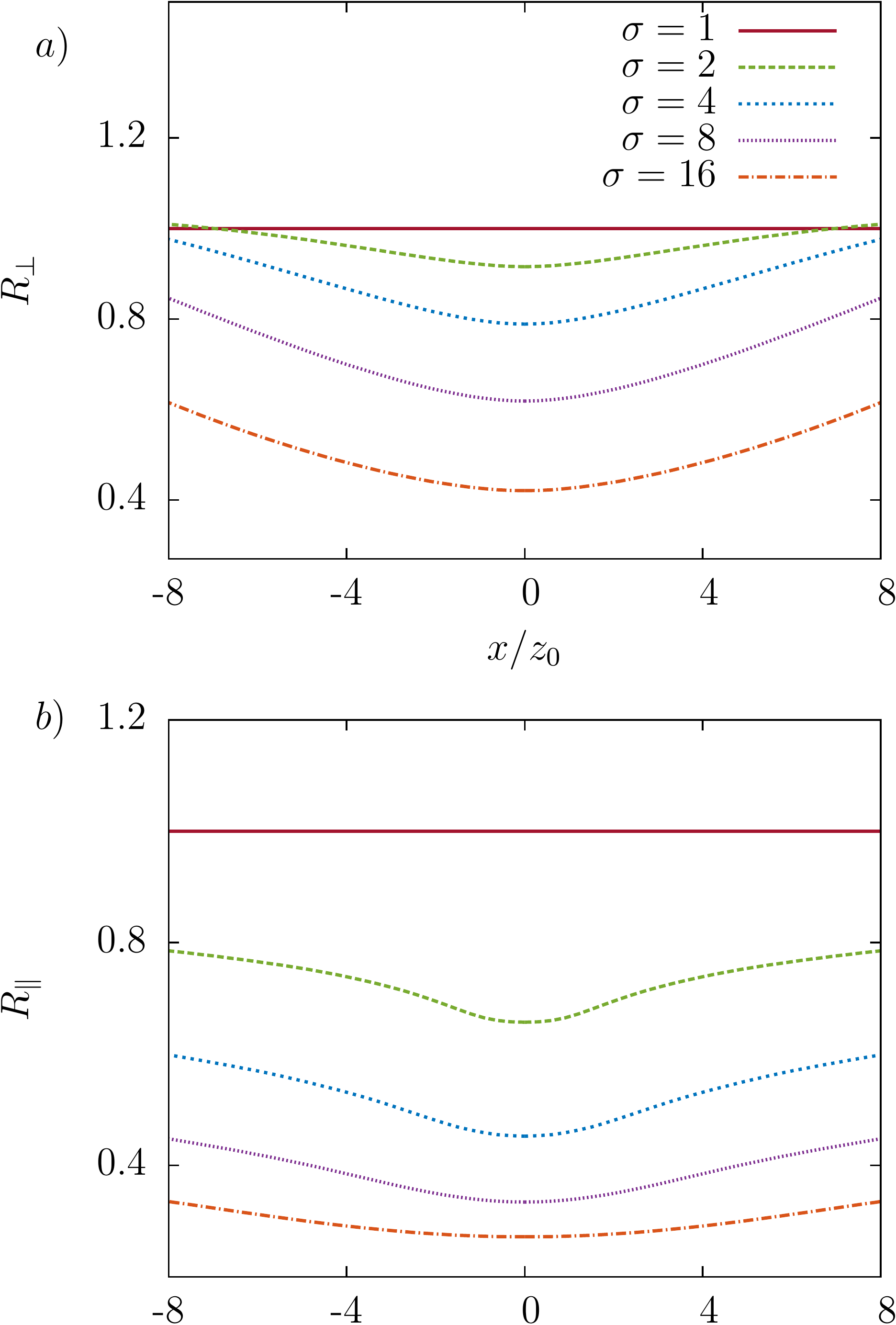}
 \caption{(Color online) $(a)$ $R_{\perp}$ and $(b)$ $R_{\parallel}$ for different values of $\sigma$ for $\omega_0\tau = 0.01$ with $z_0 = (3\kB/(2\kS))^{1/2}$.}
 \label{sigma_Effect}
 \end{center}
\end{figure}

We now consider the membrane deformation induced by the moving particle. For this, we set the complex driving force to be harmonic with components ${F}_{\alpha}(t) = A_{\alpha} e^{i\omega_0 t}$, whose temporal Fourier transform is $F_\alpha (\omega) = 2\pi A_\alpha \delta (\omega - \omega_0)$. 
In this case, the membrane displacement is  expressed as
\begin{equation}
 u_{\alpha} (\rho, \theta, t) = \psi_{\alpha \beta} (\rho, \theta, \omega_0) A_\beta e^{i\omega_0 t} \, .
 \label{displacementExpoForce}
\end{equation}
The physical displacement of the membrane is obtained by simply taking the real part of the right hand side in Eq.~\eqref{displacementExpoForce}.

Fig.~\ref{shapeModulusAll_CDL} depicts a comparison of the membrane displacements between analytical predictions and BIM simulations.
Here we use the same set of parameters as in Fig.~\ref{CDL_twoMem}.
As the particle is equally distant from both membranes, the displacement fields of each membrane are equal in magnitude, but may differ in sign.
For instance, for a particle moving perpendicularly to the membranes, the normal displacements of each membrane have the same sign whereas the radial displacements have opposite signs.
However, the vertical displacements in the parallel motion have different signs from each other whereas the in-plane displacements have similar signs.
Hereafter, all the components are evaluated in their plane of maximal displacement: $u_z$ and $u_x$  in the plane $y=0$ and $u_y$ in the plane $y=x$. 
The theoretical predictions are found to be in good agreement with the numerical simulations for both the perpendicular and parallel motions.
{The reason behind the small discrepancy between theory and simulation is most likely the fact that the analytical theory treats truly infinite membranes whereas the corresponding BIM simulations necessarily only account for finite sized membranes.}

In the perpendicular motion, the deformation is more pronounced in the normal than in the $x$-direction. The maximum displacement for the first occurs at the center.
Far away, the membrane deformation decays rapidly with distance and vanishes as $x$ tends to infinity.
On the other hand, radial symmetry implies that the displacement $u_r$ should vanish at the origin, suggesting the existence of an extremum at some intermediate radial position. The latter is found to be in magnitude around 40 times smaller than that obtained for the normal displacement. Accordingly, the in-plane deformation does not play a significant role for the motion perpendicular to the membranes.

Considering the translational motion parallel to the membranes, we observe that the displaced membranes exhibit a fundamentally different shape.
Not surprisingly, it turns out that the in-plane deformation $u_x$ along the direction parallel to the applied force is the most significant.
The maximum displacements reached in $u_z$ and $u_y$ are respectively found to be about twice and 10 times smaller in comparison with that reached in $u_x$.

Membrane deformability is largely determined by shearing and bending properties.
Henceforth, we shall consider a typical case for which both effects have the same relevance.
Thus, before we can continue, we define the characteristic time scale for shearing as $\TS := 6\eta z_0/\kS$ and the characteristic time scale for bending as $\TB := 4\eta z_0^3 / \kB$ \cite{daddi16}.
Both time scales are equal for a distance $z_0 = (3\kB/(2\kS))^{1/2}$.
We adapt this value for the remainder of this section.
Furthermore, let $\tau:= \TS = \TB$. 

It is also of interest to compute the maximum displacement (amplitude) of the membrane during the particle oscillation. The maximum is not necessarily reached for $t\omega_0 = \pi/2$, as taken in Fig.~\ref{shapeModulusAll_CDL}.
In Fig.~\ref{omega_Effect}, we show the effect of frequency on the oscillation amplitude.
Higher frequencies induce  smaller deformation, because the membrane does not have enough time to respond to the fast particle wiggling.
By comparing the reaction tensor amplitudes with and without a second membrane, we see that the presence of a second membrane reduces $|\psi_{zz}|$ less strongly than $|\psi_{xx}|$.
This is similar to the observations for the MSD in the next section (see Fig.~\ref{CDL_twoMemDiffusion}).

In order to examine the effect of the disposition of the upper membrane relative to the lower one, 
we define the following ratios of the reaction tensor amplitudes between the upper and lower membranes:
\begin{equation}
 R_{\perp} := \frac{|\psi_{zz}|_{\text{upper}}}{|\psi_{zz}|_{\text{lower}}} \quad \text{and} \quad 
 R_{\parallel} := \frac{|\psi_{xx}|_{\text{upper}}}{|\psi_{xx}|_{\text{lower}}} \, .
\end{equation}
These are two quantities that vanish for $\sigma \to \infty$ and are equal to one for $\sigma = 1$.
In Fig.~\ref{sigma_Effect}, we plot the variations of $R_{\perp}$ and $R_{\parallel}$ as functions of the scaled distance from the membrane center for different values of $\sigma$.
Here the calculations are carried out in the plane of maximal displacement $y=0$, for a scaled frequency of $\omega_0\tau = 0.01$.
We remark that the upper membrane shows significantly less vertical displacement as $\sigma$ increases (ratio less than unity.)
Further apart from the center, where less deformation occurs, the two membranes have an essentially comparable deformation behavior, 
and both ratios approach the upper limit one as $x$ increases.


\subsection{Brownian motion}

\begin{figure}
 \begin{center}
   \includegraphics[scale = 0.5]{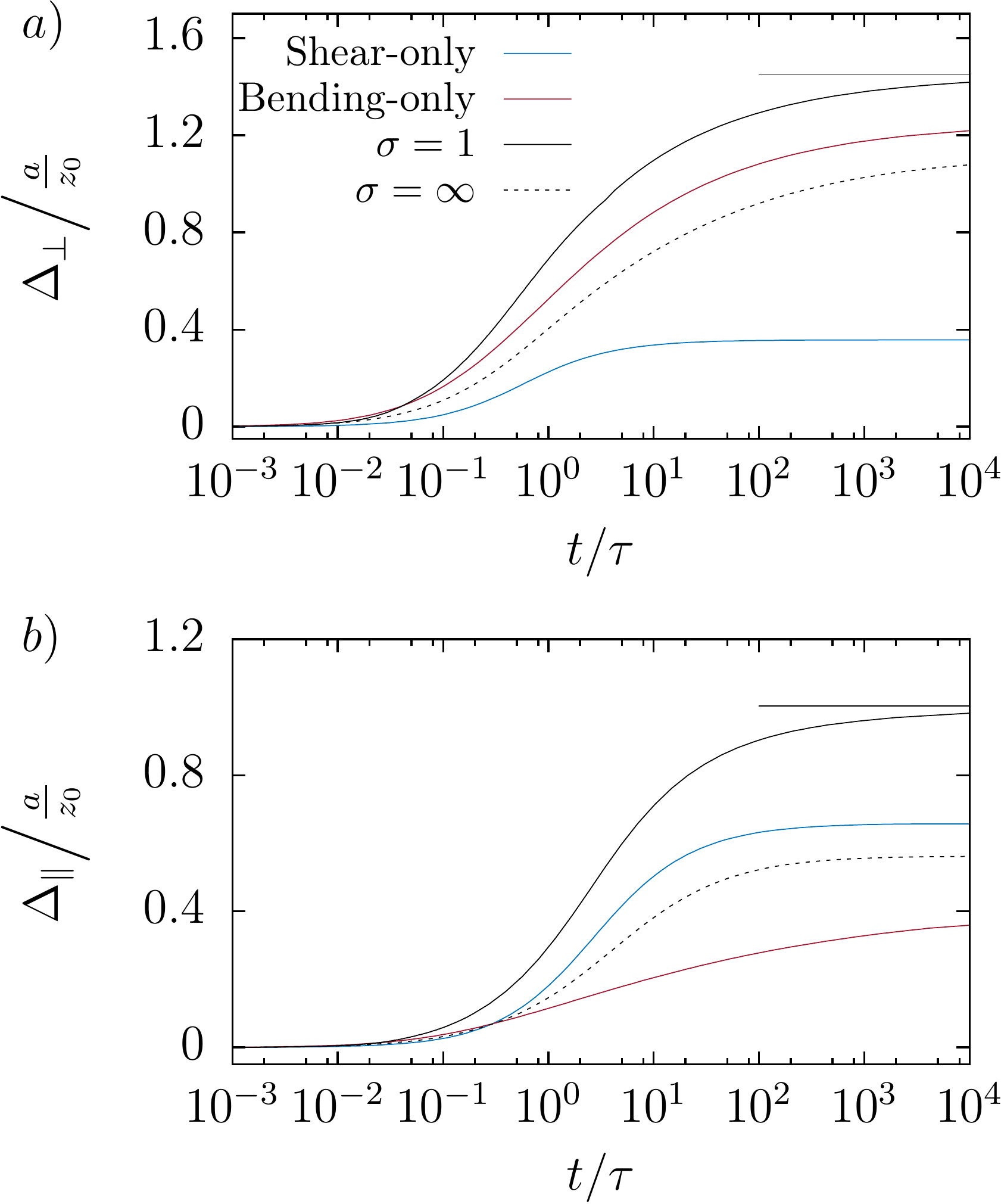}
 \caption{(Color online) Scaled excess MSD versus the scaled time for the perpendicular $(a)$ and parallel $(b)$ motions with $\sigma=1$ (black solid line) and $\sigma = \infty$ (black dotted line). A shear-only and a bending-only membrane are shown in blue and red, respectively.
 The horizontal solid line corresponds to the two-hard-walls limits.
 For the other parameters, see the main text.
 }
 \label{CDL_twoMemDiffusion} 
 \end{center}
\end{figure}

\begin{figure}
 \begin{center}
  \includegraphics[scale = 0.5]{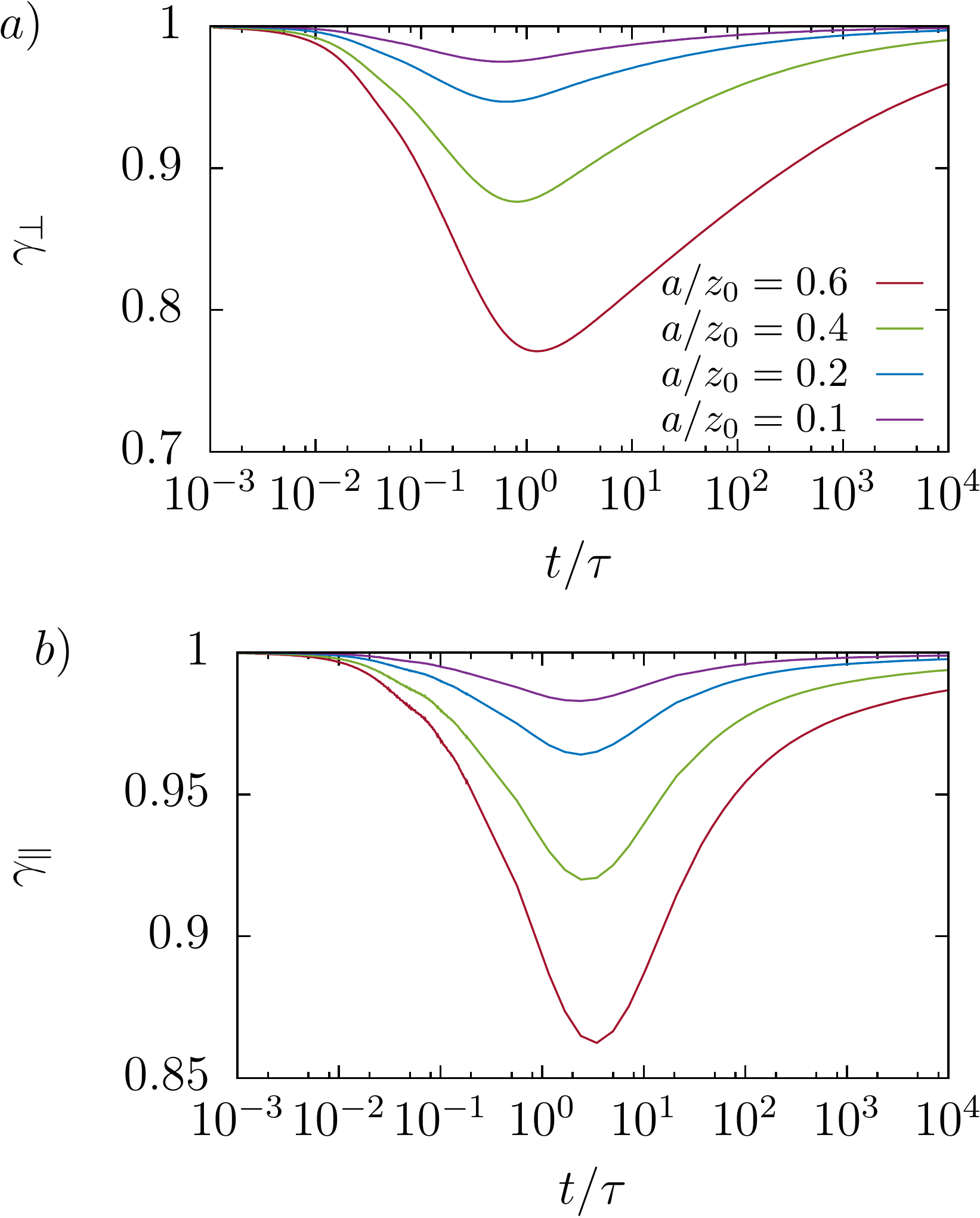}
 \caption{(Color online) Variations of the scaling exponent for the motion perpendicular $(a)$ and parallel $(b)$ to the membranes as given by Eq.~\eqref{scalingExponent} versus the scaled time for $\sigma=1$.
 }
 \label{scalingExpo} 
 \end{center}
\end{figure}

The computation of the particle mean-square displacement (MSD)  requires as an intermediate step the determination of the velocity autocorrelation function $\phi_{v,\alpha} (t) := \langle V_{\alpha}(0) V_{\alpha}(t) \rangle$.
The latter is related to the temporal inverse Fourier transform of the particle mobility via Kubo's fluctuation-dissipation theorem (FDT) 
{
such that \cite{kubo85}
\begin{equation}
 \phi_{v,\alpha} (t) = \frac{\kBolt T}{2\pi} \int_{-\infty}^{\infty} \left( \mu_{\alpha\alpha} (\omega) + \overline{\mu_{\alpha\alpha} (\omega)} \right) e^{i\omega t} \Intd \omega \, , 
\end{equation}
where $\kBolt$ is the Boltzmann constant and $T$ the absolute temperature of the system.
The bar denotes complex conjugate.}

The particle MSD is computed as
\begin{equation}
 \langle \Delta r_\alpha (t)^2 \rangle = 2 \int_0^t (t-s) \phi_{v,\alpha} (s) \, \Intd s \, .
 \label{MSD}
\end{equation}
For convenience, we define the excess MSD as
\begin{equation}
 \Delta_\alpha (t) := 1 - \frac{\langle \Delta r_\alpha (t)^2 \rangle}{2 D_0 t} \, ,
 \label{excessMSD}
\end{equation}
where $D_0 = \mu_0 \kBolt T$ is the bulk diffusion coefficient given by the Einstein relation \cite{einstein05}.

We show in Fig.~\ref{CDL_twoMemDiffusion} the variations of the perpendicular and parallel excess MSDs as computed from Eq.~\eqref{excessMSD} versus the scaled time. For short times, the particle does not yet perceive the membranes and thus experiences a bulk diffusion.
By increasing the time up to $t \approx \tau$, the effect of the confining membranes becomes noticeable. 
By comparing the total excess MSDs for $\sigma = \infty$ and $\sigma = 1$ we find that diffusion in the long-time limit is slowed down by a factor 1.78 for the parallel direction, but only a factor 1.29 in the perpendicular direction, due to the introduction of the second membrane.

As explained in Sec.~\ref{sec:coupling}, the particle mobility and, consequently, also the MSD cannot be split up directly into a shear and bending contribution for the two membrane case. We therefore consider the two cases separately, taking one membrane with $\alpha=0$ and one with $\alphaB=0$.
We find that for the shear-only membrane ($\alphaB=0$, blue curve in Fig.~\ref{CDL_twoMemDiffusion}) the time needed to reach the steady state is about $10\tau$ for the perpendicular motion, and about $100\tau$ for the parallel motion. On the other hand, the bending-only membrane ($\alpha=0$, red curve in Fig.~\ref{CDL_twoMemDiffusion}) takes for both directions a significantly longer time of about $10^4 \tau$ before the steady state is attained.

\begin{figure}
 \begin{center}
  \includegraphics[scale = 0.47]{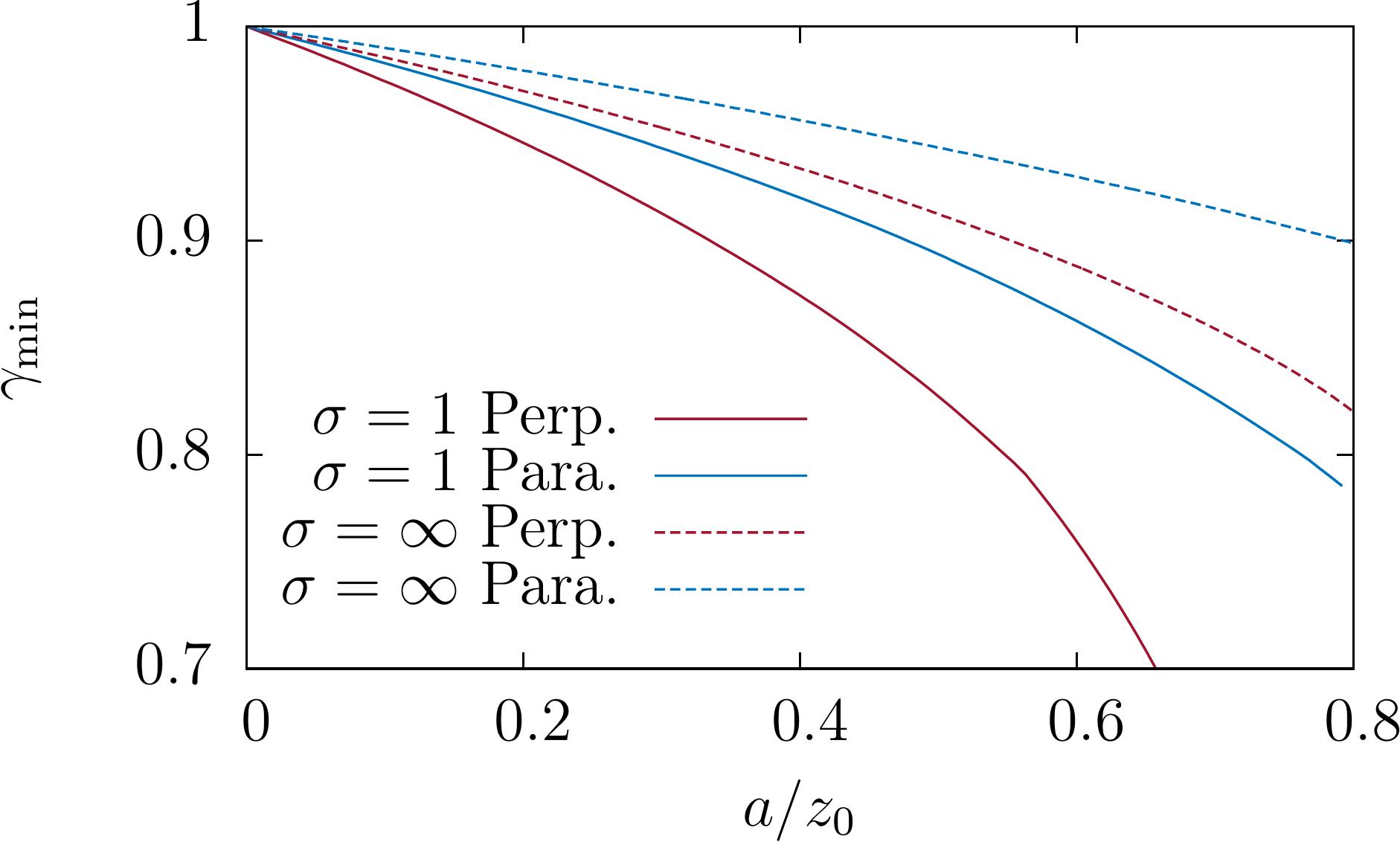}
 \caption{(Color online) Minimum of the scaling exponent versus $a/z_0$ for the perpendicular and parallel motions, for $\sigma=1$ and $\sigma=\infty$.}
 \label{minimalScalingExpo} 
 \end{center}
\end{figure}

Another way to quantify the slowing down of the particle is to investigate the time-dependent scaling exponent of the MSD, which can be defined as
\begin{equation}
\gamma_\alpha (t) := \frac{\Intd \ln \langle \Delta r_\alpha (t) ^2 \rangle}{\Intd \ln t} = 1 -\frac{t}{1-\Delta_\alpha (t)} \frac{\Intd \Delta_\alpha (t)}{\Intd t} \, .
 \label{scalingExponent} 
\end{equation}
Fig.~\ref{scalingExpo} shows the temporal evolution of the scaling exponent which strongly depends on the distance separating the particle from the membranes. 
We first remark that the scaling exponent is $\gamma (t)=1$ at $t=0$ and for $t\to\infty$. The particle thus experiences normal diffusion in these cases. This is similar to the single-membrane case \cite{daddi16}.
For $t \approx \tau$, we observe a bending down of the scaling exponent, resulting in a subdiffusive regime that extends up to 10$^3 \tau$ in the parallel and even further in the perpendicular direction.
In Fig.~\ref{minimalScalingExpo} we present the variation of the minimal scaling exponent for $\sigma = 1$ and $\sigma=\infty$ upon varying the particle-membrane distance.
For $a/z_0=0.6$, the exponent is found to be as low as 0.75 for the perpendicular motion, and 0.86 for the parallel motion.
These values are significantly smaller than the ones previously found in the one-membrane limit ($\sigma = \infty$) \cite{daddi16}, where the scaling exponent is around 0.89 and 0.92 for the perpendicular and parallel motions, respectively.
We therefore conclude that the second membrane leads to a notable slow-down of the dynamics.


\section{Conclusions}
\label{conclusions}

We have investigated the translational motion of a spherical particle confined between two parallel elastic membranes and determined the frequency dependent mobility for the motion perpendicular and parallel to the membranes in the point particle limit.
Contrary to the single wall, shear and bending are intrinsically coupled and their contributions cannot be added linearly.
Our analytical predictions have been compared to boundary integral simulations for a finite-sized particle and very good agreement has been observed.
The frequently used superposition approximation, originally suggested by Oseen \cite{oseen28} for two hard walls, has been tested for elastic membranes. 
Reasonably good agreement with the analytically exact predictions is observed for the parallel, but not for the perpendicular motion, especially in the low frequency regime.

Subsequently, we have provided analytical predictions validated by numerical simulations of the membrane deformation due to a particle upon which an oscillating force is exerted perpendicular or parallel to the membranes.
We have observed that the deformation is most pronounced in the direction along which the force acts, and that the presence of the second membrane significantly reduces the membrane deformations.

Finally, we have shown that the elastic membranes induce a memory effect in the system, leading to a subdiffusive Brownian motion at intermediate time scales. This is qualitatively similar, yet more pronounced, as in the single membrane situation \cite{daddi16}.
To provide typical physical values, consider a red blood cell with a shear modulus of $\kS = 5 \times 10^{-6} \, \mathrm{N}/\mathrm{m}$ and a bending modulus of $\kB = 2\times 10^{-19} \, \mathrm{N m}$ that flows in a fluid with dynamic viscosity $\eta = 1.2 \times 10^{-3} \, \mathrm{Pa} \, \mathrm{s}$ \cite{freund13}.
{A typical nanoparticle of radius $a = 150 \, \mathrm{nm}$ that is located at a distance of $z_0 = 250 \, \mathrm{nm}$ from both red blood cells will undergo a long-lived subdiffusive motion that can last up to $100 \, \mathrm{ms}$. }
The corresponding scaling exponent of the MSD can go as low as 0.77 in the perpendicular and as low as 0.87 in the parallel direction.

In the future, it will be interesting to carry out similar calculations in more severe confinements such as cylindrical elastic channels where even stronger effects are expected.

\begin{acknowledgements}
The authors gratefully acknowledge funding from the Volkswagen Foundation and the KONWIHR network as well as computing time granted by the Leibniz-Rechenzentrum on SuperMUC.
\end{acknowledgements}

\appendix*

\section{Computation of the traction jump for the membranes}
\label{appendixForceComputations}

In this appendix, we provide some technical details regarding the computation of the traction jump $\Delta \vect{f}$ across the membranes, as required for Eq.~\eqref{functionH_BIM}.
The membranes are endowed with shear and area elasticity together with some bending rigidity.

\subsection{Shear and area elasticity}
We employ the Skalak model \cite{skalak73} which is often used to model the membranes of red blood cells.
Its areal energy density is given by \cite{kruger11}
\begin{equation}
	\epsilon_\mathrm{S} = \frac{\kS}{12} (I_1^2 + 2 I_1 - 2 I_2 + C I_2^2) \, .
\end{equation}

The strain invariants $I_1$ and $I_2$ are related to the principal in-plane stretch ratios via $I_1 = \lambda_1^2 + \lambda_2^2 - 2$ and $I_2 = \lambda_1^2 \lambda_2^2 - 1$.
Hence, the total energy of a membrane $S_{\mathrm{m}_i}$ is given by
\begin{equation}
	E_\mathrm{S} = \int_{S_{\mathrm{m}_i}^{(0)}} \epsilon_\mathrm{S} \, \Intd S_0 \, ,
\end{equation}
where the integration is performed over the surface in the reference state $S_{\mathrm{m}_i}^{(0)}$. In our case this is a simple flat sheet.
To obtain the force at each node, we assume that the deformation is a linear function of position in each triangle. After discretization of the integral the energy $E_\mathrm{S}$ depends explicitly on the node positions $\vect{x}_i$. Therefore, according to the principle of virtual work, the total force is then given by the gradient
\begin{equation}
	\vect{F}(\vect{x}_i) = \frac{\partial E_\mathrm{S}}{\partial \vect{x}_i} \, .
\end{equation}
This derivative can be computed analytically as detailed in references~\cite{Kraus1996,Kruger2012}.
The traction jump is thus obtained by
\begin{equation}
	\Delta \vect{f}(\vect{x}_i) = \frac{\vect{F}(\vect{x}_i)}{A_i} \, ,
\end{equation}
whereas $A_i$ is the area associated with node $\vect{x}_i$ and is taken as one third of the total area of the triangles containing the node \cite{Zhao2012a}.

\subsection{Bending rigidity}
The bending forces are modeled according to the constitutive law proposed by Canham \cite{Canham1970} and Helfrich \cite{helfrich73}, which for a flat reference state becomes
\begin{equation}
	E_\mathrm{B} = 2 \kB \int_{S_{\mathrm{m}_i}} H^2 \, \Intd  S \, .
\end{equation}
$H$ denotes the mean curvature and $\kB$ the bending modulus.
Applying the principle of virtual work is possible \emph{before} the discretization, leading to the following contribution to the traction jump \cite{Zhong-Can1989a,Laadhari2010}:
\begin{equation}
	\Delta \vect{f}(\vect{x}) = - 2 \kB \left(2H(H^2 - K) + \Delta_\mathrm{S} H\right)\vect{n} \, .
\end{equation}
The mean curvature $H$ is calculated according to the relation $H(\vect{x}) = -\frac{1}{2}\left(\Delta_\mathrm{S} x_i \right) n_i(\vect{x})$.
We use the algorithms presented by Meyer \textit{et al.}~\cite{Meyer2002a} for the computation of the Laplace-Beltrami operator $\Delta_\mathrm{S}$ and the Gaussian curvature $K$.
The normal vector $\vect{n}$ is computed according to the ``{mean weighted by angle}'' method \cite{Jin2005}.
This provides reasonable results in the application of viscous flows \cite{guckenberger16}.
Note that we set $\Delta \vect{f}$ to zero for nodes located at the border of the meshes.


\input{main.bbl}

\end{document}

%% file: main.bbl
%